\documentclass{jpsj2}
\usepackage{epsf}
\usepackage{graphicx}
\usepackage{multirow}
\usepackage{booktabs}
\usepackage{amsfonts}
\usepackage{dcolumn}%
\usepackage{color,ulem}
\usepackage{bm}

\begin{document}
 
\title{Magnetization Reversal by Tuning Rashba Spin-Orbit Interaction and Josephson Phase in a Ferromagnetic Josephson Junction} 

\author{Shin-ichi Hikino}
\inst{%
National Institute of Technology, Fukui College, Sabae, Fukui 916-8507, Japan
} 

\date{\today}

\abst{
We theoretically investigate the magnetization inside a normal metal containing the Rashba spin-orbit interaction (RSOI) 
induced by the proximity effect in an {\it s}-wave superconductor/normal metal
/ferromagnetic metal/{\it s}-wave superconductor ({\it S/N/F/S}) Josephson junction. 
By solving the linearized Usadel equation taking account of the RSOI, 
we find that the direction of the magnetization induced by the proximity effect in {\it N} can be reversed by tuning the RSOI. 
Moreover, we also find that the direction of the magnetization inside {\it N} can be reversed by changing the superconducting phase difference, i.e., Josephson phase. 
From these results, it is expected that the dependence of the magnetization on the RSOI and Josephson phase can be applied to superconducting spintronics. 
}


\maketitle 


\section{Introduction}\label{sec:introduction}
In {\it s}-wave superconductor/ferromagnetic metal ({\it S/F}) junctions, 
it is well known that the pair amplitude of a spin-singlet Cooper pair (SSC) penetrating into {\it F} due to the proximity effect shows damped oscillatory 
behavior as a function of the thickness of {\it F}~\cite{ryazanov-prl,kontos-prl,sellier-pr,bauer-prl,
frolov-prb,robinson,born-prb,weides-prl,oboznov-prl,shelukhin-prb,pfeiffer-prb,bannykh-prb,khaire-prb,wild-epjb,kemmer-prb,yamashita-pra8,
golubov-rmp,buzdin-rmp,bergeret-rmp}. 
One of the interesting phenomena resulting from the damped oscillatory behavior of the pair amplitude is a $\pi$-state in 
an {\it S}/{\it F}/{\it S} junction, where the current--phase relation in the Josephson current is shifted by $\pi$ from that of 
ordinary {\it S}/{\it I}/{\it S} and {\it S}/{\it N}/{\it S} junctions~\cite{ryazanov-prl,kontos-prl,sellier-pr,bauer-prl,
frolov-prb,robinson,born-prb,weides-prl,oboznov-prl,shelukhin-prb,pfeiffer-prb,bannykh-prb,khaire-prb,wild-epjb,kemmer-prb,yamashita-pra8,
golubov-rmp,buzdin-rmp,bergeret-rmp}. 

Another notable phenomenon in {\it S/F} junctions is the appearance of odd-frequency spin-triplet Cooper pairs (STCs) 
induced by the proximity effect\cite{bergeret-rmp,linder-arxiv}. 
In an {\it S}/{\it F} junction having a uniform magnetization, 
the STC composed of opposite spin electrons ($S_z=0$) and the SSC penetrate into {\it F} owing to the 
proximity effect~\cite{bergeret-rmp,yokoyama-prb75}. 
The penetration length of the STC with $S_z=0$ and the SSC into {\it F} is determined by $\xi_{\rm F}=\sqrt{\hbar D_{\rm F}/h_{\rm ex}}$, which is typically 
of nm order~\cite{ryazanov-prl, kontos-prl, sellier-pr, bauer-prl, frolov-prb, robinson, born-prb, weides-prl, oboznov-prl, shelukhin-prb, pfeiffer-prb, bannykh-prb, khaire-prb, wild-epjb, kemmer-prb,yamashita-pra8, golubov-rmp, buzdin-rmp, bergeret-rmp}. 
Here, $D_{\rm F}$ and $h_{\rm ex}$ are the diffusion coefficient and exchange field in {\it F}, respectively. 

Moreover, the STC formed by electrons of equal spin ($|S_z|=1$) can be induced inside {\it F} owing to the proximity effect 
when the magnetization in {\it F} is nonuniform in {\it S/F} junctions. 
The penetration length of the STC with spin $|S_{z}|=1$ is determined by $\xi_{\rm T}=\sqrt{\hbar D_{\rm F}/2 \pi k_{\rm B} T}$ in $F$ 
({\it T} is temperature). 
The feature of this STC is approximately two orders of magnitude larger than the penetration length of the SSC and STC with $S_{z}=0$~\cite{bergeret-prl86, champel-prb72, braude-prl98, fominov-prb75, 
volkov-prb78, alidoust-prb81, bzdin-prb83,volkov-prb90, bergeret-prb68, 
houzet-prb76, trifunovic-prb82, volkov-prb81, trifunovic-prb84, melnikov-prl109,knezevic-prb85, richard-prl110, fritsch-njp16, alidoust-prb89, 
fominov-jetpl77, fominov-jetpl91, kawabata-jpsj82, mironov-prb89, halterman-prb94, eschrig-sh, asano-prl, galaktionov-prb, beri-prb, linder-prb82, trifunovic-prl107, bergeret-prl110, pal-sr7}. 
Thus, the proximity effect of the STC with $|S_{z}|=1$ is referred to as the long-range proximity effect (LRPE). 

The STC with $|S_{z}|=1$ induced by the proximity effect can be detected by the observation of Josephson current in ferromagnetic Josephson junctions (FJJs). 
The Josephson current carried by the STC with $|S_{z}|=1$ monotonically decreases as a function of the thickness of {\it F} and the 
decay length of the STC is roughly determined by $\xi_{\rm T}$. 
The long-range Josephson current flowing through the {\it F} has been observed and established experimentally 
in FJJs\cite{keizer,robinson-science,khaire,anwar-apl,leksin-prl109,wang-prl89,singh-prx5}. 
The detection of long-range Josephson current is a piece of evidence for the presence of the STC. 

Another means of obtaining evidence for the presence of the STC is to measure the spin-dependent transport of the STC in {\it S/F} junctions, 
since such transport can be used to directly measure the spin of the STC~\cite{Lofwander-prl95,halterman-prb77,shomali-njp,pugach-apl101,kukagna-prb90,moor-sst28,hikino-prb92,hikino-jpsj86}. 
For this purpose, {\it S/F/N} and {\it S/F/N/F/S} junctions containing the Rashba spin-orbit interaction (RSOI) have been of considerable interest in recent years, 
since the proximity effect coupled with the RSOI in {\it S/F/N} and {\it S/F/N/F/S} junctions exhibits many fascinating phenomena which are not observed in 
{\it S/F/N} and {\it S/F/N/F/S} junctions without the RSOI~\cite{liu-apl96, bergeret-su2, liu-prl113, kon-prb92, alid-njp17, jacb-scr6, linder-prb96}. 
For {\it S/F/N} junctions, it has been theoretically found 
that the pair amplitude of the STC penetrating into {\it N} with the RSOI is modulated as a function of the thickness of {\it N} and the magnitude of the RSOI~
\cite{bergeret-su2, liu-prl113}. 
By utilizing the modulation of the pair amplitude of the STC, 
it is possible to freely control 0- and $\pi$-states by tuning the RSOI in Josephson junctions~\cite{liu-apl96,liu-prl113}. 
Moreover, some authors have theoretically predicted that the direct evidence of the STC can also be obtained 
by detecting the spin Hall effect and magnetoelectric effect induced by the STC 
in Josephson junctions containing ferromagnetic metals and the RSOI~\cite{kon-prb92,alid-njp17,linder-prb96}. 
The RSOI is  advantageous for studying spin-dependent transport phenomena since it can be freely controlled by an external electric field~\cite{sb-book}. 
However, understanding of the spin transport of the STC in consideration of the RSOI is still lacking, 
since the studies in this research field have been rare and limited~\cite{kon-prb92,alid-njp17,linder-prb96}. 
Therefore, it is expected that a good understanding of the spin transport of the STC will provide proof for the STC and 
expedite the development of superconducting spintronics~\cite{esch-ss,linder-np,jac-sci, y-apl110}. 

In this paper, we theoretically propose another setup and way to detect the STC 
by using an {\it S}1/{\it N}/{\it F}/{\it S}2 junction differently from Refs. 71, 72, and 74. 
The {\it S}1/{\it N}/{\it F}/{\it S}2 junction can be easily achieved by using recent device fabrication techniques. 
Based on the linearized Usadel equation including the RSOI, 
we formulate the magnetization induced by the proximity effect. 
It is shown that the magnetization in the {\it N} only appears when the product of the anomalous Green's functions of 
the spin-triplet odd-frequency Cooper pair and spin-singlet even-frequency Cooper pair in the {\it N} has a finite value. 
It is found that the magnetization shows damped oscillatory behavior as a function of the thickness of {\it N}. 
It is also found that the direction of magnetization can be controlled by tuning the magnitude of the RSOI. 
Moreover, we examine the Josephson phase ($\theta$) dependence of the magnetization. 
It is found that the period of oscillation of magnetization can be changed by tuning $\theta$. 
This result clearly shows that the direction of the magnetization can also be controlled by tuning $\theta$ as well as the magnitude of the RSOI. 
Therefore, we expect that these results can be applied to the research field of superconducting spintronics. 

The rest of this paper is organized as follows. 
In Sect.~\ref{sec:formulation}, we introduce an  {\it S}1/{\it N}/{\it F}/{\it S}2 Josephson junction and 
formulate the magnetization induced by the proximity effect in the $N$ containing the RSOI of this junction by solving the Usadel equation. 
In Sect.~\ref{sec:results}, the numerical results of the magnetization are given. 
The thickness and RSOI dependences of the magnetization are discussed. 
Moreover, we present the Josephson phase ($\theta$) dependence of the magnetization. 
Finally, the $\theta$-magnetization relation is discussed and 
the magnetization induced by the proximity effect is estimated for a typical set of realistic parameters in Sect.~\ref{sec:dis}.
A summary of this paper is given in Sect.~\ref{sec:sum}. 
The detailed calculation of the magnetization is given in the Appendices.

\section{Magnetization in Normal Metal Induced by Proximity Effect in an S1/N/F/S2 Junction with Rashba Spin-Orbit Interaction}\label{sec:formulation}

\begin{figure}[t!]
\begin{center}
\includegraphics[width=7cm]{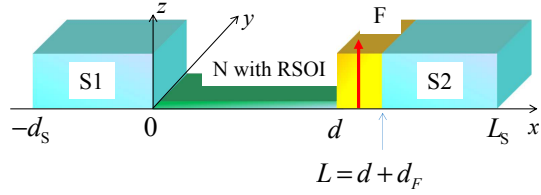}
\caption{ (Color online) 
Schematic illustration of the $S$1/$N$/$F$/$S$2 junction studied, where $N$ is a normal metal with the RSOI, 
$F$ is a ferromagnetic metal, and {\it S}1(2) is an $s$-wave superconductor. 
The arrow in $F$ indicates the direction of the ferromagnetic magnetization 
where the magnetization in $F$ is fixed along the $z$ direction. 
$d_{\rm S}$, $d_{\rm F}$, and $d$ are the thicknesses of $S$, $F$, and $N$, 
respectively, with $L=d+d_{\rm F}$. 
We assume that the magnetization is uniform in the $F$ layer and that  
$d_{\rm S}$ is much larger than $\xi_{\rm S}$. 
}
\label{s2dfs-gm}
\end{center}
\end{figure}

\subsection{Setup of junction and anomalous Green's functions}\label{sec:agf}
We consider a Josephson junction composed of $s$-wave superconductors ($Ss$) 
separated by a normal metal/ferromagnetic metal ($N/F$) junction as depicted in Fig.~\ref{s2dfs-gm}. 
Here, we include the RSOI in the $N$ and assume uniform magnetization in $F$ in the $S1/N/F/S2$ junction. 
We assume that the width of the junction is smaller than $\xi_{\rm T}$. 
In this situation, a one-dimensional (1D) model may be a good approximation. 
Therefore, we adopt the 1D model to analyze the magnetization induced by the proximity effect in the $S1/N/F/S2$ junction. 

In the diffusive transport limit, the magnetization inside the $N$ with the RSOI induced by the proximity effect is 
evaluated by solving the linearized Usadel equation including the SU(2) gauge field in each region ($j=N, F$),\cite{bergeret-su2} 
\begin{equation}
i \hbar D_{j} \tilde{\partial }^{2}_{x} \hat{f}^{j}(\bm{r}) 
-i2\hbar |\omega_{n}| \hat{f}^{j}(\bm{r}) 
-{\rm sgn}(\omega_{n}) h_{\rm ex}(x) 
\left[
\hat{\tau}_{z}, \hat{f}^{j}(\bm{r}) 
\right]
= \hat{0} 
\label{usadel},  
\end{equation}
\begin{eqnarray}
{{\tilde \partial }_x} = \left\{ \begin{array}{l}
{\partial _x} \bullet  - i\frac{1}{\hbar }\left[ {{{\hat A}_x}, \bullet } \right],\,\,\,\,\,\,{{\hat A}_x} = {\alpha _R}{{\hat \tau }_y},\,\,\,0 < x < d\\
{\partial _x} \bullet \,\,\,\,\,\,\,\,\,\,\,\,\,\,\,\,\,\,\,\,\,\,\,\,\,\,\,\,\,\,\,\,\,\,\,,\,\,\,{\rm{otherwise}}
\end{array} \right., \nonumber
\end{eqnarray}
where ${\bm r}=(x,\omega_{n})$, $\hat{A}_{x} $ is the SU(2) gauge field, which describes the RSOI, and $\alpha_{\rm R}$ is the RSOI constant. 
$D_{j}$ is the diffusion coefficient in region $j$, 
$\omega_n=(2n+1)\pi k_{\rm B}T/\hbar$ ({\it n}: integer) is the fermion Matsubara frequency, 
${\rm sgn}(X)=X/|X|$ is the sign function, and $\hat{\tau}_{y (z)}$ is the $y(z)$ component of the Pauli matrix.
$[\hat{Q}, \hat{R}]=\hat{Q}\hat{R}-\hat{R}\hat{Q}$ is the anticommutation relation and $\hat{ 0}$ is the zero matrix. 
The anomalous part $\hat{f}^{j}(\bm{r})$ of the quasiclassical Green's function \cite{eschrig-sh} is given by 
\begin{eqnarray}
{\hat f^j}\left( \bm{r} \right) &=& \left( {\begin{array}{*{20}{c}}
	{f_{ \uparrow  \uparrow }^j\left( \bm{r} \right)}&{f_{ \uparrow  \downarrow }^j\left( \bm{r} \right)}\\
	{f_{ \downarrow  \uparrow }^j\left( \bm{r} \right)}&{f_{ \downarrow  \downarrow }^j\left( \bm{r} \right)}
	\end{array}} \right) \nonumber \\
	&=& \left( {\begin{array}{*{20}{c}}
	{ - f_{tx}^j\left( \bm{r} \right) + if_{ty}^j\left( \bm{r} \right)}&{f_s^j\left( \bm{r} \right) + f_{tz}^j\left( \bm{r} \right)}\\
	{ - f_s^j\left( \bm{r} \right) + f_{tz}^m\left( \bm{r} \right)}&{f_{tx}^j\left( \bm{r} \right) + if_{ty}^j\left( \bm{r} \right)}
\end{array}} \right)
\label{f}. 
\end{eqnarray}
$f_s^j(\bm{r})$ is the anomalous Green's function for the SSC, and $f_{tx(ty)}^j(\bm{r})$ and $f_{tz}^j(\bm{r})$ represent the anomalous Green's functions for 
the STC with $|S_z|=1$ and $|S_z|=0$, respectively. 
The exchange field ${\bm h}_{\rm ex}(x)$ in the {\it F} is given by 
\begin{eqnarray}
{\bm h_{{\rm{ex}}}}\left( x \right) = \left\{ \begin{array}{l}
{h_{{\rm{ex}}}}{{\bm e}_z}\,,\,\,\,\,d < x < L\\
0,\,\,\,\,\,\,{\rm{otherwise}}
\end{array} \right.,
\end{eqnarray}
where $\bm{e}_{z}$ is a unit vector in the $z$ direction. 
We assume that $h_{\rm ex}$ is positive. 

To obtain solutions of Eq.~(\ref{usadel}), we employ appropriate boundary conditions, i.e., 
\begin{eqnarray}
{\left. {{{\hat f}^{{\rm{S1}}}(\bm{r}) }} \right|_{x = 0}} &=& {\left. {{{\hat f }^{{\rm{N}}}}(\bm{r}) } \right|_{x = 0}} 
\label{bc1}, \\
{\left. {{{\hat f}^{\rm{N}}(\bm{r})}} \right|_{x = d}} &=& {\left. {{{\hat f}^{{\rm{F}}}(\bm{r}) }} \right|_{x = d}}
\label{bc2}, \\
{\left. {{{\hat f}^{{\rm{F}}}(\bm{r}) }} \right|_{x = L}} &=& {\left. {{{\hat f}^{{\rm{S2}}}(\bm{r}) }} \right|_{x = L}} 
\label{bc3}, \\
\sigma_{\rm S} {\left. {\partial _x}{{\hat f}^{S1}}\left( \bm{r} \right) \right|_{x=0}} &=& \sigma_{\rm N} {\left. {\partial _x}{{\hat f}^{\rm N}}\left( \bm{r} \right) \right|_{x=0}}, 
\label{bc4} \\
{\left. {\partial_x {{\hat f}^{{\rm{F}}}(\bm{r}) }} \right|_{x = {d}}} &=& {\left. \frac{1}{{{\gamma _{\rm{F}}}}} {\partial_x {{\hat f}^{\rm{N}}(\bm{r}) }} \right|_{x = {d}}}
\label{bc5}, \\
\sigma_{\rm S} {\left. {\partial _x}{{\hat f}^{S2}}\left( \bm{r} \right) \right|_{x=L}} &=& \sigma_{\rm F} {\left. {\partial _x}{{\hat f}^F}\left( \bm{r} \right) \right|_{x=L}} 
\label{bc6}, 
\end{eqnarray}
where $\gamma_{\rm F}=\sigma_{\rm F}/\sigma_{\rm N}$ and 
$\sigma_{\rm F(N)}$ is the conductivity of $F$($N$). 
Moreover, in the present calculation,  we adopt the rigid boundary condition 
$\frac{\sigma_{\rm F(N)}}{\sigma_{\rm S}} \ll \frac{\xi_{\rm F(N)}}{\xi_{\rm S}}$,
where $\sigma_{\rm S}$ is the conductivity of $S$ in the normal state~\cite{buzdin-rmp}. 
$\xi_{\rm F}=\sqrt{\hbar D_{\rm F}/h_{\rm ex }}$ and $\xi_{\rm N}=\sqrt{\hbar D_{\rm N}/2\pi k_{\rm B} T}$.  
Notice that the left-hand side of Eqs.~(\ref{bc4}) and (\ref{bc6}) is zero as will be shown later. 
Assuming that $d_{\rm S}\gg\xi_{\rm S}$, the anomalous Green's function in the {\it S}s attached to {\it N} and {\it F} 
can be approximately given as 
\begin{equation}
{\hat f}_{s}^{\rm S1(2)} (\bm{r}) = - {\hat \tau}_{y} 
\frac{\Delta_{\rm L (R)}  } {\sqrt{(\hbar \omega)^{2} + |\Delta_{\rm L (R)}|^{2}}}\equiv \hat{F}^{\rm S1(2)}  
\label{fs}, 
\end{equation} 
where $\Delta_{\rm L(R)} = \Delta e^{i \theta_{\rm L(R)}}$ ($\Delta$: real) and $\theta_{\rm L(R)}$ 
is the superconducting phase in the left (right) side of the $S$ (see Fig.~\ref{s2dfs-gm}). 
The {\it s}-wave superconducting gap $\hat{\Delta}(x)$ is finite only in the {\it S} and is assumed to be constant as follows: 
\begin{eqnarray}
\hat \Delta(x)  = \left\{ \begin{array}{l}
\left( {\begin{array}{*{20}{c}}
0&{ - \Delta_{\rm L} }\\
\Delta_{\rm L} &0
\end{array}} \right), - {d_{\rm S}} < x < 0 \\
\left( {\begin{array}{*{20}{c}}
0&{ - \Delta_{\rm R} }\\
\Delta_{\rm R} & 0
\end{array}} \right), L < x<L_{\rm S} \\ 
\,\,\,\,\,\,\,\,\,\,\,\,\,\,\,\,\,\,\hat 0,\,\,\,\,\,\,\,\,\,\,\,\,\,\,\,\,\,\,\,\,\,\,\,\,\,\,\,\,\,\,{\rm{otherwise}}
\end{array} \right.\ .
\label{gap}
\end{eqnarray}
From Eqs.~(\ref{fs}) and (\ref{gap}), it is immediately found that the left-hand side of Eqs.~(\ref{bc4}) and (\ref{bc6}) becomes zero, 
since the rigid boundary condition is assumed in the present calculation\cite{buzdin-rmp}. 

Assuming $d_{\rm F}/\xi_{\rm F}\ll 1$, we can perform the Taylor expansion with $x$ for $\hat{f}^{\rm F}(\bm{r}) $ as follows:\cite{houzet-prb76, ode-book} 
\begin{eqnarray}
{\hat f^{\rm F}}\left( \bm{r} \right) &\approx& {\hat f^{\rm F}}\left( d,\omega_{n} \right) 
	+ \left. \left( {x - d} \right){\partial _x}{\hat f^{\rm F}}\left( \bm{r} \right) \right|_{x=d} \nonumber \\
	& + & \frac{{{{\left( {x - d} \right)}^2}}}{2}\left. \partial _x^2{\hat f^{\rm F}}\left( \bm{r} \right)\right|_{x=d}
\label{ff}.
\end{eqnarray}
Applying the boundary conditions of Eqs.~(\ref{bc2}) and (\ref{bc5}) to Eq.~(\ref{ff}) and substituting Eq.~(\ref{ff}) into Eq.~(\ref{usadel}), 
we can approximately obtain the anomalous Green's function of $f^{\rm F}(\bm{r})$ as 
%
%
\begin{eqnarray}
{\hat f}^{\rm F}(\bm{r}) &\approx &
		-\frac{d_{\rm F}}{\gamma_{\rm F}} \left. \partial _{x} \hat{f}^{\rm N}(\bm{r})\right|_{x=d}
		+\frac{(x-d)}{\gamma_{\rm F}} \left. \partial _{x} \hat{f}^{\rm N}(\bm{r})\right|_{x=d}
		+ {\hat F}^{\rm S2}
		+ i {\rm sgn}(\omega_{n}) \frac{h_{\rm ex} d_{\rm F}^{2} }{2 \hbar D_{\rm F}}
		\left[ 
		\hat{\tau_{z}}, {\hat F}^{\rm S2}
		\right] \nonumber \\
		&-& i {\rm sgn}(\omega_{n}) \frac{(x-d)^{2} h_{\rm ex}}{2\hbar D_{\rm F}} 
		\left[
		\hat{\tau_{z}}, {\hat F}^{\rm S2}
		\right]
\label{ff1x}.
\end{eqnarray}
%

The general solutions of Eq.~(\ref{usadel}) in the $N$ are given by
\begin{eqnarray}
\left[ \begin{array}{l}
f_s^{\rm{N}}\left( {\bm r} \right)\\
f_{tx}^{\rm{N}}\left( {\bm r} \right)\\
f_{tz}^{\rm{N}}\left( {\bm r} \right)
\end{array} \right] &=& \left[ \begin{array}{l}
{A_1}\\
0\\
0
\end{array} \right]{e^{{k_{\rm{N}}}x}} + \left[ \begin{array}{l}
{A_2}\\
0\\
0
\end{array} \right]{e^{ - {k_{\rm{N}}}x}}, \nonumber \\
&+& B\left[ \begin{array}{l}
0\\
i\\
1
\end{array} \right]{e^{i\tilde \alpha x}}{e^{{k_\alpha }x}} + C\left[ \begin{array}{l}
0\\
i\\
1
\end{array} \right]{e^{i\tilde \alpha x}}{e^{ - {k_\alpha }x}} + F\left[ \begin{array}{l}
0\\
i\\
1
\end{array} \right]{e^{ - i\tilde \alpha x}}{e^{{k_\alpha }x}} + G\left[ \begin{array}{l}
0\\
i\\
1
\end{array} \right]{e^{ - i\tilde \alpha x}}{e^{ - {k_\alpha }x}}, 
\label{fn1}
\end{eqnarray}
where $k_{\alpha} = \sqrt{3 \alpha_{\rm R}^{2} + k_{\rm N}^{2}}$ and $\sqrt{2|\omega_{n}|/D_{\rm N}}$. 
Here, we assume that $\alpha_{\rm R}\not =0$ to obtain Eq.~(\ref{fn1}). 
Applying the boundary conditions given in Eqs.~(\ref{bc1}), (\ref{bc3}), (\ref{bc4}), and (\ref{bc6}) to Eq.~(\ref{fn1}) 
, and also using the result in Eq.~(\ref{ff}), we can obtain the anomalous Green's functions in the $N$ as 
%
\begin{eqnarray}
f^{\rm N}_{s} (\bm{r}) &=& 
\left[
-i \frac{\Delta_{\rm L}}{E_{\omega_{n}}} 
\left(
1-\frac{k_{\rm N} d_{\rm F}}{\gamma_{\rm F}}
\right)
\sinh[k_{\rm N}(x-d)] 
+
i \frac{\Delta_{\rm R}}{E_{\omega_{n}}} 
\sinh(k_{\rm N}x)
\right] 
Q_{\omega_{n}}(d), 
\label{fns} \\
f^{\rm N}_{tx}(\bm{r}) &=& i f^{\rm N}_{tz}(\bm{r}), 
\label{fntx}
\end{eqnarray}
%
and
\begin{eqnarray}
f^{\rm N}_{tz}(\bm{r}) &=& {\rm sgn} (\omega_{n}) \frac{h_{\rm ex}d_{\rm F}^{2}}{\hbar D_{\rm F}} \frac{\Delta_{\rm R}}{E_{\omega_{n}}}
\Phi_{\omega_{n}}(d) t_{\omega_{n}}(x,d), 
\label{fntz}
\end{eqnarray}
where $E_{\omega_{n}}=\sqrt{(\hbar \omega_{n})^{2} + \Delta^{2}}$, 
\begin{equation}
Q_{\omega_{n}}^{-1}(d) = 
\sinh(k_{\rm N}d) + \frac{k_{\rm N}d_{\rm F}}{\gamma_{\rm F}} \cosh(k_{\rm N}d), 
\label{qwn}
\end{equation}
%
\begin{eqnarray}
\Phi_{\omega_{n}}^{-1}(d) &=&
\left( {i{\alpha _{\rm{R}}} + {k_\alpha }} \right)\frac{{{d_{\rm{F}}}}}{{{\gamma _{\rm{F}}}}}\left[ {{e^{\left( {i{\alpha _{\rm{R}}} + {k_\alpha }} \right)L}} 
+ {e^{ - \left( {i{\alpha _{\rm{R}}} + {k_\alpha }} \right)L}}} \right]{C^{31}_{\omega_{n}}}\left( d \right) \nonumber \\
&+& \frac{{{d_{\rm{F}}}}}{{{\gamma _{\rm{F}}}}}\left[ {\left( {i{\alpha _{\rm{R}}} 
- {k_\alpha }} \right){e^{\left( {i{\alpha _{\rm{R}}} - {k_\alpha }} \right)L}} 
+ \left( {i{\alpha _{\rm{R}}} + {k_\alpha }} \right){e^{ - \left( {i{\alpha _{\rm{R}}} 
+ {k_\alpha }} \right)L}}} \right]{C^{32}_{\omega_{n}}}\left( d \right) \nonumber \\
&-& \frac{{{d_{\rm{F}}}}}{{{\gamma _{\rm{F}}}}}\left[ {\left( {i{\alpha _{\rm{R}}} - {k_\alpha }} \right){e^{\left( { - i{\alpha _{\rm{R}}} 
+ {k_\alpha }} \right)L}} - \left( {i{\alpha _{\rm{R}}} + {k_\alpha }} \right)
{e^{ - \left( {i{\alpha _{\rm{R}}} + {k_\alpha }} \right)L}}} \right]{C^{33}_{\omega_{n}}}\left( d \right)
\label{phiw},
\end{eqnarray}
and
%
\begin{eqnarray}
{t_{{\omega _n}}}\left( {x,d} \right) &=& 
	i2\left[ {C_{{\omega _n}}^{21}\left( d \right) 
	+ 2C_{{\omega _n}}^{31}\left( d \right)} \right] \nonumber \\
	&\times & \left[ {\cos \left( {{\alpha _{\rm{R}}}d} \right)\sinh \left( {{k_\alpha }x} \right) 
	+ i\sin \left( {{\alpha _{\rm{R}}}d} \right)\cosh \left( {{k_\alpha }x} \right)} \right] \nonumber \\
	&-& 2\left[ {C_{{\omega _n}}^{22}\left( d \right) + 2C_{{\omega _n}}^{32}\left( d \right)} \right]\sin \left( {{\alpha _{\rm{R}}}x} \right){e^{ - {k_\alpha }x}} \nonumber \\ 
	&+& i2\left[ {C_{{\omega _n}}^{23}\left( d \right) + 2C_{{\omega _n}}^{33}\left( d \right)} \right]\sinh \left( {{k_\alpha }x} \right){e^{ - i{\alpha _{\rm{R}}}x}}
\label{txw}, 
\end{eqnarray}
%
where the explicit formulae of the functions $C_{\omega_{n}}^{i j}(d)$ $(i,j=1,2,3)$ are presented in Appendix A. 
From Eqs.~(\ref{fns})--(\ref{fntz}), it is immediately found that 
$f_{s}^{\rm N}(\bm r)$ representing the SSC is an even function with $\omega_{n}$, 
whereas $f_{tx(tz)}^{\rm N}(\bm{r})$ representing the STC is an odd function with $\omega_{n}$ 
since $f_{tx(tz)}^{\rm N}(\bm{r})$ is proportional to ${\rm sgn}(\omega_{n})$. 
Therefore, $f_{tx(tz)}^{\rm N}(\bm{r})$ describes the odd-frequency STC. 
Note that $f_{ty}^{\rm N}(\bm{r}) = 0$ since we assume that the present junction is a 1D model and the magnetization in the {\it F} has only a {\it z} component~\cite{bergeret-su2}. 
Also note that $f_{tx(tz)}^{\rm N}(\bm{r})$ is exactly zero when $h_{\rm ex}=0$, which corresponds to no magnetic layer in the present junction studied here. 
This result shows that the RSOI does not induce the STC in the $N$, and thus the magnetic layer is needed to produce the STC. 
The role of the RSOI is to induce a finite $f_{tx}^{\rm N}(\bm{r}) $ in the $N$ of the present junction\cite{note1}.

\subsection{Magnetization induced by proximity effect in normal metal with RSOI}
On the basis of the quasiclassical Green's function theory, 
the magnetization $\bm{M}(d,\theta)$ induced by the proximity effect is given by\cite{champel-prb72, Lofwander-prl95}
\begin{eqnarray}
	\bm{M}(d,\theta) &=&
		(M_{x}(d,\theta),M_{y}(d,\theta),M_{z}(d,\theta)) \nonumber \\
		&=&
		\frac{A}{V}
		\int_{0}^{d}
		\bm{m}(x,\theta) dx
\label{md},
\end{eqnarray}
where $\theta=\theta_{\rm R}-\theta_{\rm L}$ is the Josephson phase in the junction and 
\begin{eqnarray}
	\bm{m} (x,\theta) &=& (m_{x}(x,\theta),m_{y}(x,\theta),m_{z}(x,\theta)) \nonumber \\
	&=&- g \mu_{\rm B} \pi N_{\rm F} k_{\rm B} T 
	\sum_{\omega_{n}}
	{\rm sgn}(\omega_{n})
	{\rm Im}
	\left[
	f_{s}^{\rm N}(\bm{r}) \bm{f}_{t}^{\,\,\rm N*}(\bm{r}) 
	\right] \nonumber \\
\label{lm}
\end{eqnarray}
with 
\begin{equation}
{\bm{f}_{t}}^{\,\,\rm N}(\bm{r}) = 
	(f_{tx}^{\rm N}(\bm{r}),-f_{ty}^{\rm N}(\bm{r}),f_{tz}^{\rm N}(\bm{r}))
\label{vec-f}.
\end{equation}
$\bm m(x,\theta)$ is the local magnetization density 
in the $N$, $g$ is the $g$ factor of an electron, and $\mu_{\rm B}$ is the Bohr magneton. 
$A$ and $V=Ad$ are the cross-section area of the junction and the volume of $N$, respectively. 
$N_{\rm F}$ is the density of states per unit volume and per electron spin at the Fermi energy. 

It is obvious from Eq.~(\ref{lm}) that  $f_{s}^{\rm N}({\bm r})$ and $\bm{f}_{t}^{\,\,\rm N}({\bm r})$ must both be nonzero to induce a finite 
$\bm{m} (x,\theta)$. However, as described in Sect.~\ref{sec:agf}, a nonzero $\bm{f}_{t}^{\,\,\rm N}({\bm r})$ 
occurs only when the $F$ layer is involved in the junction. 
Therefore, the origin of the magnetization in the $N$ is considered to be the STCs induced 
by the proximity effect~\cite{bergeret-rmp,Lofwander-prl95,pugach-apl101, hikino-prb92}.  
Because $f_{ty}^{\rm N}({\bm r})=0$ (see Sect.~\ref{sec:agf}), $m_{y}(x,\theta)$ and $M_{y} (d,\theta)$ are always zero. 
It is also noticeable that $M_{x}^{(2)}(d,\theta)$ and $M_{x}^{(3)}(d,\theta)$ are 
only induced in the $N$ when the RSOI and the Josephson coupling are finite in the {\it S/N/F/S} junction. 
This result is in sharp contrast to Josephson junctions composed of a metallic multilayer system without the RSOI\cite{pugach-apl101, hikino-prb92, hikino-jpsj86}.
Therefore, in what follows, we only consider the $x$ and $z$ components of $\bm{M}(d,\theta)$. 

Substituting Eq.~(\ref{fns}) and the complex conjugate of Eq.~(\ref{vec-f}) into Eq.~(\ref{lm}), 
and integrating Eq.~(\ref{lm}) with respect to $x$ from 0 to $d$, we can obtain the $x$ and $z$ components of the magnetization given by Eq.~(\ref{md}). 
The $x$ component $M_{x}(d,\theta)$ is decomposed into three parts, 
\begin{eqnarray}
M_{x}(d,\theta) &=&
M_{x}^{(1)}(d) + M_{x}^{(2)}(d,\theta) + M_{x}^{(3)}(d,\theta), 
\label{mdx}
\end{eqnarray}
where
%
\begin{eqnarray}
M_{x}^{(1)}(d) &=&
- g \mu_{\rm B} \pi N_{\rm F} k_{\rm B} T
\frac{h_{\rm ex} d_{\rm F}^{2}}{\hbar D_{\rm F}d} \nonumber \\
	&\times& \sum_{\omega_{n}} 
\frac{\Delta^{2}}{E_{\omega_{n}}^{2}}
Q_{\omega_{n}}(d)
{\rm Im}
\left[
\Phi_{\omega_{n}}(d) w_{\omega_{n}}(d)
\right]
\label{mxd1}, \\
M_{x}^{(2)}(d,\theta) &=&
g \mu_{\rm B} \pi N_{\rm F} k_{\rm B} T
\frac{h_{\rm ex} d_{\rm F}^{2}}{\hbar D_{\rm F}d} 
\sum_{\omega_{n}} 
\frac{\Delta^{2}}{E_{\omega_{n}}^{2}}
\left(
1-\frac{k_{\rm N}d_{\rm F}}{\gamma_{\rm F}}
\right) \nonumber \\
	&\times& Q_{\omega_{n}}(d)
{\rm Im}
\left[
\Phi_{\omega_{n}}(d) u_{\omega_{n}}(d)
\right] \cos\theta
\label{mxd2}, 
\end{eqnarray}
%
and 
\begin{eqnarray}
M_{x}^{(3)}(d,\theta) &=&
g \mu_{\rm B} \pi N_{\rm F} k_{\rm B} T
\frac{h_{\rm ex} d_{\rm F}^{2}}{\hbar D_{\rm F}d} 
\sum_{\omega_{n}} 
\frac{\Delta^{2}}{E_{\omega_{n}}^{2}}
\left(
1-\frac{k_{\rm N}d_{\rm F}}{\gamma_{\rm F}}
\right) \nonumber \\
	&\times& Q_{\omega_{n}}(d)
{\rm Re}
\left[
\Phi_{\omega_{n}}(d) u_{\omega_{n}}(d)
\right] \sin\theta 
\label{mxd3}. 
\end{eqnarray}
%
Similarly, the $z$ component $M_{z}(d,\theta)$ is also decomposed into three parts, 
\begin{eqnarray}
M_{z}(d,\theta) &=&
M_{z}^{(1)}(d) + M_{z}^{(2)}(d,\theta) + M_{z}^{(3)}(d,\theta), 
\label{mdz}
\end{eqnarray}
where 
%
\begin{eqnarray}
M_{z}^{(1)}(d) &=&
g \mu_{\rm B} \pi N_{\rm F} k_{\rm B} T
\frac{h_{\rm ex} d_{\rm F}^{2}}{\hbar D_{\rm F}d} \nonumber \\
	&\times& \sum_{\omega_{n}} 
\frac{\Delta^{2}}{E_{\omega_{n}}^{2}}
Q_{\omega_{n}}(d)
{\rm Re}
\left[
\Phi_{\omega_{n}}(d) w_{\omega_{n}}(d)
\right]
\label{mzd1}, \\
M_{z}^{(2)}(d,\theta) &=&
- g \mu_{\rm B} \pi N_{\rm F} k_{\rm B} T
\frac{h_{\rm ex} d_{\rm F}^{2}}{\hbar D_{\rm F}d} 
\sum_{\omega_{n}} 
\frac{\Delta^{2}}{E_{\omega_{n}}^{2}}
\left(
1-\frac{k_{\rm N}d_{\rm F}}{\gamma_{\rm F}}
\right) \nonumber \\
	&\times& Q_{\omega_{n}}(d)
{\rm Re}
\left[
\Phi_{\omega_{n}}(d) u_{\omega_{n}}(d)
\right] \cos \theta
\label{mzd2}, 
\end{eqnarray}
%
and
\begin{eqnarray}
M_{z}^{(3)}(d,\theta) &=&
g \mu_{\rm B} \pi N_{\rm F} k_{\rm B} T
\frac{h_{\rm ex} d_{\rm F}^{2}}{\hbar D_{\rm F}d} 
\sum_{\omega_{n}} 
\frac{\Delta^{2}}{E_{\omega_{n}}^{2}}
\left(
1-\frac{k_{\rm N}d_{\rm F}}{\gamma_{\rm F}}
\right) \nonumber \\
	&\times& Q_{\omega_{n}}(d)
{\rm Im}
\left[
\Phi_{\omega_{n}}(d) u_{\omega_{n}}(d)
\right] \sin\theta 
\label{mzd3}. 
\end{eqnarray}
%
The explicit formulae of the functions $w_{\omega_{n}}(d)$ and $u_{\omega_{n}}(d)$ in Eqs.~(\ref{mxd1})--(\ref{mzd3}) are given in Appendix B. 
From Eqs.~(\ref{mxd1})--(\ref{mxd3}) and Eqs.~(\ref{mzd1})--(\ref{mzd3}), 
it is immediately found that the magnetizations of the $x$ and $z$ components are exactly zero when the exchange field $h_{\rm ex}$ is zero. 
Therefore, the $F$ is indeed required to induce the magnetization inside the $N$. 
Note that $M_{x}(d,\theta)$ is always zero without the RSOI as mentioned above.

One of the $\theta$-independent parts of the magnetization, i.e., $M_{z}^{(1)}(d)$, 
is due to the proximity effect common in the $S/F$ multilayer systems\cite{bergeret-rmp,pugach-apl101, hikino-prb92, hikino-jpsj86}. 
The other  $\theta$-independent part of the magnetization, i.e., $M_{x}^{(1)}(d)$, is due to not only the proximity effect but also the presence of the RSOI, 
since the magnetization of the $F$ is oriented along the $z$ axis in the present junction\cite{bergeret-rmp}. 
The $\theta$-dependent part of $M_{z}^{(2)}(d,\theta)$ is induced by the coupling between the two superconductors 
only when $S/F$ multilayer systems compose the Josephson junction\cite{pugach-apl101, hikino-prb92, hikino-jpsj86}. 
The $\theta$-dependent part of $M_{x}^{(2)}(d,\theta)$ is induced by the coupling between the two superconductors and the finite RSOI. 
$M_{x}^{(3)}(d,\theta)$ and $M_{z}^{(3)}(d,\theta)$ appear 
when the RSOI, the exchange field, and the Josephson coupling are finite. 
The expressions for the magnetizations $M_{x}(d,\theta)$ and $M_{z}(d,\theta)$ given in Eqs.~(\ref{mdx})--(\ref{mzd3}) are rather complicated. 
Therefore, we will present numerical results of magnetizations calculated here in the next section.

\begin{figure}[b!]
\begin{center}
\vspace{20 mm} 
\includegraphics[width=8cm]{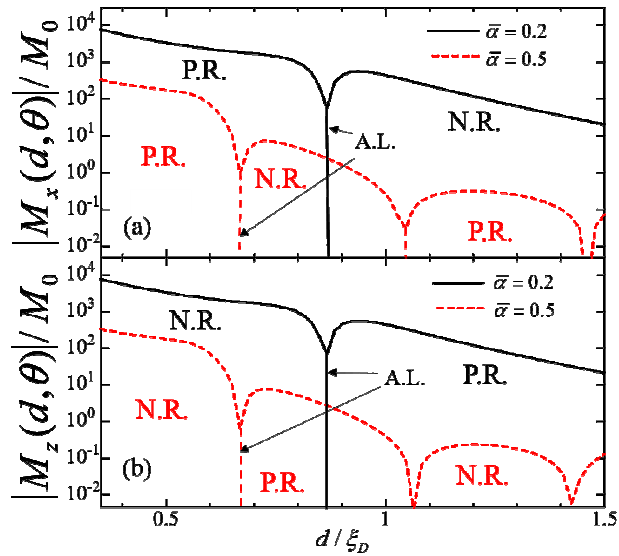}
\caption{(Color online) 
(a) $x$ component $M_{x}(d,\theta)$ and (b) $z$ component $M_{z}(d,\theta)$ of magnetization 
in the $N$ as a function of $d$ for ${\bar{\alpha} }$ = 0.2 and 0.5. 
$M_{x}(d,\theta)$ and $M_{z}(d,\theta)$ show the damped oscillatory behavior with $d$, 
where P.M and N.R. are, respectively, positive and negative regions of $M_{x}(d,\theta)$ and $M_{z}(d,\theta)$. 
Here we set $T/T_{\rm C} =0.5$, $\gamma_{\rm F}=0.1$, $\theta=\pi/4$, $d_{\rm F}/\xi_{\rm D}=0.01$, and $h_{\rm ex}=30$. 
$\bar{\alpha}=\alpha_{\rm R}\xi_{\rm D} $ and $\xi_{\rm D}=\sqrt{\hbar D_{\rm N}/2\pi k_{\rm B}T_{\rm C}}$. 
}
\label{md1}
\end{center}
\end{figure}

\begin{figure}[t!]
\begin{center}
\includegraphics[width=7.5cm]{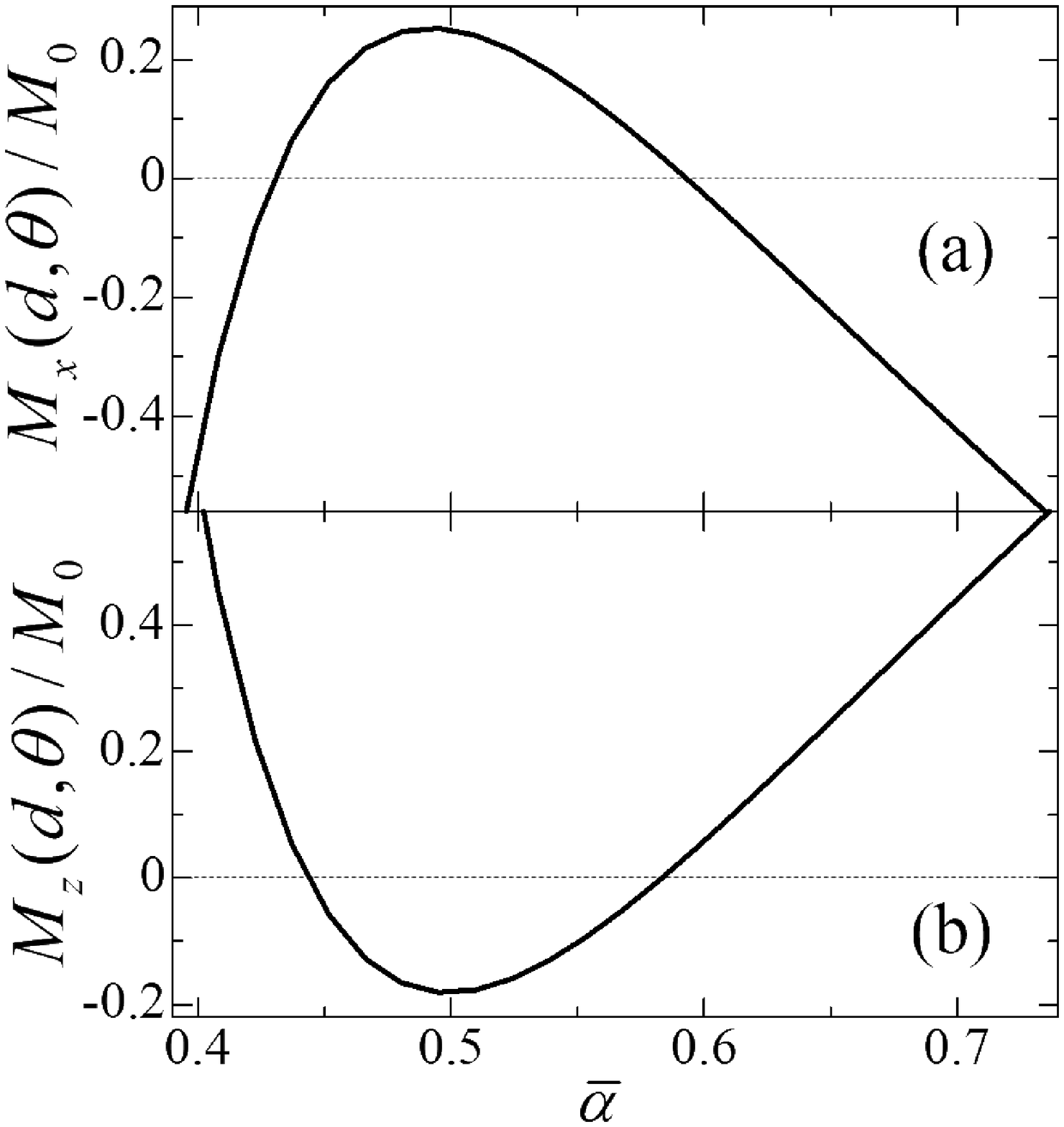}
\caption{(Color online) 
(a) $x$ component $M_{x}(d,\theta)$ and (b) $z$ component $M_{z}(d,\theta)$ of magnetization 
in the {\it N} as a function of $\bar{\alpha}$ for $d/\xi_{\rm D}=1.3$. 
Here we set $T/T_{\rm C} =0.5$, $\gamma_{\rm F}=0.1$, $d_{\rm F}/\xi_{\rm D}=0.01$, and $h_{\rm ex}=30$. 
$\bar{\alpha}=\alpha_{\rm R}\xi_{\rm D} $ and $\xi_{\rm D}=\sqrt{\hbar D_{\rm N}/2\pi k_{\rm B}T_{\rm C}}$.
It is clearly found that magnetizations are reversed (a) from negative to positive values then from positive to negative values 
and (b) from positive to negative values then from negative to positive values by increasing $\bar{\alpha}$. 
}
\label{ma}
\end{center}
\end{figure}

\section{Results}\label{sec:results}
In this section, we numerically evaluate the magnetizations of Eqs.~(\ref{mdx}) and (\ref{mdz}) induced by the proximity effect in the $S/N/F/S$ junction. 
In order to perform the numerical calculation of $M_{x}(d,\theta)$ and $M_{z}(d,\theta)$, 
the temperature dependence of $\Delta$ is assumed to be $\Delta=\Delta_{0} \tanh(1.74\sqrt{T_{\rm C}/T-1})$, 
where $\Delta_{0}$ is the superconducting gap at zero temperature and $T_{\rm C}$ is the superconducting transition temperature. 
The thicknesses of $N$ and $F$ are normalized by $\xi_{\rm D}=\sqrt{\hbar D_{\rm N}/2\pi k_{\rm B} T_{\rm C}}$ and 
the magnetizations of the $x$ and $z$ components are normalized by $M_{0}=(g \mu_{\rm B} N_{\rm F} \Delta_{0})$.

\subsection{Thickness dependence of magnetizations induced by the proximity effect}
Figure~2 shows the $x$ and $z$ components of the magnetization induced by the proximity effect inside the $N$ as a function of $d$. 
The solid (black) and dashed (red) lines are magnetizations for $\bar{\alpha} =0.2$, and 0.5, respectively. 
In Fig.~\ref{md1}, P.R. and N.R. are abbreviations for positive and negative regions, respectively. 
A.L. denotes the auxiliary line separating the positive and negative regions of magnetization in Fig.~\ref{md1}. 
We find that $M_{x}(d,\theta)$ and $M_{z}(d,\theta)$ exhibit damped oscillatory behavior as a function of $d$. 
From Fig.~\ref{md1}, it is found that the magnetizations can be reversed by changing the thickness of $N$. 
Moreover, it is also clearly found that the period of oscillation of $M_{x}(d,\theta)$ and $M_{z}(d,\theta)$ becomes short with increasing $\alpha_{\rm R}$. 
Therefore, by setting $d$ near the thickness for which $M_{x(z)}(d,\theta)\approx 0$, the magnetizations can be easily reversed by tuning $\alpha_{\rm R}$.

\begin{figure}[t!]
\begin{center}
\includegraphics[width=8cm]{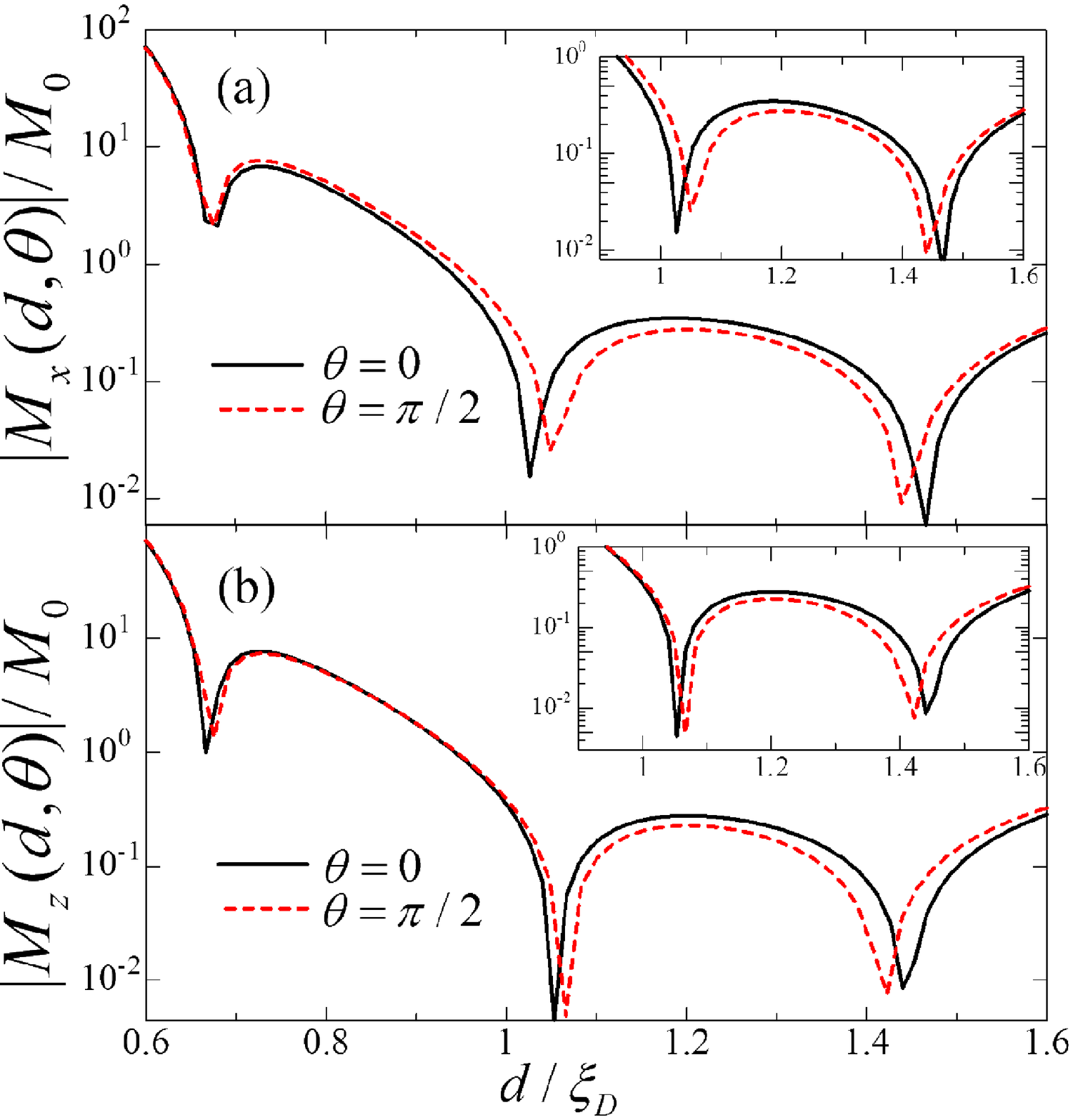}
\caption{(Color online) 
(a) $x$ component $M_{x}(d,\theta)$ and (b) $z$ component $M_{z}(d,\theta)$ of magnetization 
in the {\it N} as a function of $d$ for $\theta$ = 0 and $\pi/2$. 
Here we set $\bar{\alpha}=0.5 $, $T/T_{\rm C} =0.5$, $\gamma_{\rm F}=0.1$, $d_{\rm F}/\xi_{\rm D}=0.01$, and $h_{\rm ex}=30$. 
$\bar{\alpha}=\alpha_{\rm R}\xi_{\rm D} $ and $\xi_{\rm D}=\sqrt{\hbar D_{\rm N}/2\pi k_{\rm B}T_{\rm C}}$. 
The insets show the behavior of magnetizations from $d/\xi_{\rm D}$= 0.9 to $d/\xi_{\rm D}$=1.6. 
It is clearly found that the period of oscillation in $M_{x}(d,\theta)$ and $M_{z}(d,\theta)$ can be controlled by $\theta$. 
}
\label{md2}
\end{center}
\end{figure}

\begin{figure}[h!]
\begin{center}
\includegraphics[width=7.5cm]{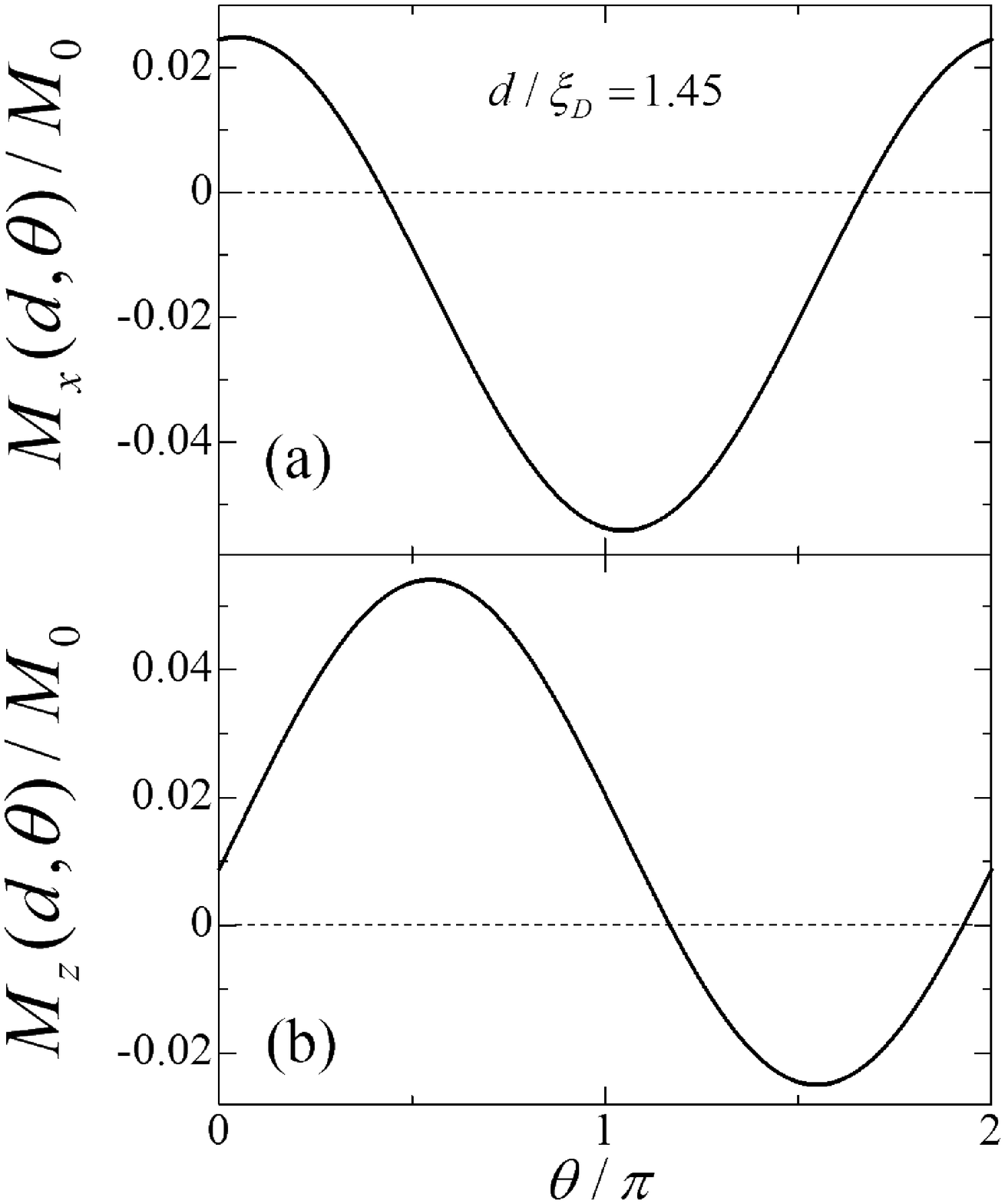}
\caption{(Color online) 
(a) $x$ component $M_{x}(d,\theta)$ and (b) $z$ component $M_{z}(d,\theta)$ of magnetization 
in the {\it N} as a function of $\theta$ for $d/\xi_{\rm D}=1.45$. 
Here we set $\bar{\alpha}=0.5 $, $T/T_{\rm C} =0.5$, $\gamma_{\rm F}=0.1$, $d_{\rm F}/\xi_{\rm D}=0.01$, and $h_{\rm ex}=30$. 
$\bar{\alpha}=\alpha_{\rm R}\xi_{\rm D} $ and $\xi_{\rm D}=\sqrt{\hbar D_{\rm N}/2\pi k_{\rm B}T_{\rm C}}$.
For $d/\xi_{\rm D}=1.45$, the magnetizations are reversed from positive to negative then from negative to positive by increasing $\theta$. 
}
\label{mq}
\end{center}
\end{figure}

Figures~\ref{ma}(a) and \ref{ma}(b) show the $x$ and $z$ components of the magnetization as a function of $\alpha_{\rm R}$, respectively. 
Here, we set the thickness of $N$ as $d/\xi_{\rm D}=1.3$. 
From Fig.~\ref{ma}(a), it is found that the sign of $M_{x}(d,\theta)$ is changed from negative to positive with increasing $\alpha_{\rm R}$ 
within $0.4\xi_{\rm D} \lesssim \alpha_{\rm R} \lesssim 0.6\xi_{\rm D}$. 
The sign of $M_{x}(d,\theta)$ is then changed from positive to negative with further increasing $\alpha_{\rm R}$. 
From Fig.~\ref{ma}(b), it is found that the sign of $M_{z}(d,\theta)$ is changed from negative to positive with increasing $\alpha_{\rm R}$ 
within $0.4\xi_{\rm D} \lesssim \alpha_{\rm R} \lesssim 0.6\xi_{\rm D}$. 
By further increasing $\alpha_{\rm R}$, the sign of $M_{z}(d,\theta)$ is changed from positive to negative. 
From these results, it is clearly found that the direction of the $x$ and $z$ components of the magnetization can be reversed by tuning $\alpha_{\rm R}$.

\subsection{RSOI dependence of magnetization and magnetization-phase relation}
Figure~\ref{md2} shows the magnetization as a function of $d$. 
Figures~\ref{md2}(a) and \ref{md2}(b) show the $x$ and $z$ components of the magnetization, respectively. 
In Fig.~\ref{md2}, the solid (black) and dashed (red) lines are the magnetizations for $\theta =0$ and $\pi/2$, respectively. 
$\bar{\alpha}$ is set to 0.5. 
From Fig.~\ref{md2}, it is found that the periods of oscillation of $M_{x}(d,\theta)$ and $M_{z}(d,\theta)$ can be changed by tuning $\theta$. 
Note that the variation of the oscillation period of the magnetizations becomes large when $d/\xi_{\rm D} >1 $ as shown in Fig.~\ref{md2}. 
By setting $d/\xi_{\rm D}$ near the third minimum of $M_{x(z)}(d,\theta)$, 
we can perform magnetization reversal by changing $\theta$ as well as $\alpha_{\rm R}$ (see Figs.~\ref{ma} and \ref{md2}). 

Figure~\ref{mq} shows the magnetization as a function of $\theta$, i.e., the magnetization--phase relation\cite{note}. 
Figures~\ref{md2}(a) and \ref{md2}(b) show the $x$ and $z$ components of the magnetization, respectively. 
The thickness of $N$ is set to $d/\xi_{\rm D}=1.45$. 
The behavior of $M_{x}(d,\theta)$ with $\theta$ is a cosine function as shown in Fig.~\ref{mq}(a). 
On the other hand, the behavior of $M_{z}(d,\theta)$ with $\theta$ is a sine function as shown in Fig.~\ref{mq}(b). 
From Fig.~\ref{mq}, it is immediately found that $M_{x}(d,\theta)$ and $M_{z}(d,\theta)$ can be varied from positive to negative values and vice versa by changing $\theta$. 
This result indicates that the magnetizations can be reversed by changing $\theta$ 
when the thickness of {\it N} is appropriately set as mentioned above. 

\section{Discussion}\label{sec:dis}
Here, we discuss why the magnetization--phase relation of $M_{x}(d,\theta)$ is shifted by $\pi/2$ compared with that of $M_{z}(d,\theta)$. 
From Eq.~(\ref{lm}), the $x$ component of the magnetization is proportional to ${\rm Im}[f_{s}(\bm r) f_{tx}(\bm r)]$. 
From Eq.~(\ref{fntx}), $f_{tx}(\bm r) = i f_{tz}(\bm r)$, ${\rm Im}[f_{s}(\bm r) f_{tx}(\bm r)] = {\rm Im}[f_{s}(\bm r) f_{tz}(\bm r) e^{i \pi/2}]$. 
Therefore, the magnetization--phase relation of $M_{x}(d,\theta)$ is shifted by $\pi/2$ compared with that of $M_{z}(d,\theta)$. 

We approximately estimate the amplitude of the magnetization induced by the proximity effect. 
As shown in Figs.~\ref{md1} and \ref{ma}, the magnetization in the $N$ has a finite value in the length scale of $\xi_{\rm D}$. 
In the dirty limit, $\xi_{\rm D}$ is on the orders of 10--100 nm\cite{book-deutscher}. 
Therefore, the magnetization induced by the proximity effect has a finite value in this length scale. 
To estimate the amplitude of the magnetization, 
we evaluate the normalized factor of the magnetization, i.e., $M_{0}=g \mu_{\rm B} N_{\rm F} \Delta_{0}$ (for instance, see Fig.~\ref{md1}). 
When we use a typical set of parameters\cite{note7, book-ssp}, $M_{0}$ is approximately 100 A/m. 
Therefore, the order of the magnetization amplitude is between $10$ and $10^{4}$ (see Fig.~\ref{md2}). 
It is expected that this order of  magnetization amplitude can be detected by magnetization measurement utilizing a SQUID\cite{book-mm}. 

Finally, we approximately estimate the thickness ($d$) of {\it N}, the thickness ($d_{\rm F}$) of {\it F}, and the magnitude of $\alpha_{\rm R}$. 
We estimate {\it d} and $d_{\rm F}$ by considering realistic materials. 
When an InGaAs/InAlAs quantum well as a normal metal and CuNi as a ferromagnetic metal are chosen, 
from Fig.~\ref{md1} and Refs. 5 and 86, the suitable $d$ and $d_{\rm F}$ are of 100 nm and nm orders, respectively. 
The total thickness between the two superconductors is of 100 nm orders. 
For this thickness, the Josephson coupling is still finite and thus the magnetization induced by proximity effect studied here 
can be controlled by the Josephson phase.
In the present calculation, we chose $\alpha_{\rm R}$ to be one order smaller than $\xi_{\rm D}$. 
The magnitude of $\alpha_{\rm R}$ used in the numerical calculation is easily achieved by performing realistic experiments
\cite{nitta-prl78,grundler-prl84,heida-prb57,manchon-nt14,shekhter-prl116}. 
For instance, InGaAs/InAlAs and InAs/AlSb quantum wells are good candidates as the {\it N} in the present junction studied here\cite{nitta-prl78,grundler-prl84,heida-prb57}. 
Therefore, it is expected that the magnetization induced by the proximity effect can be easily reversed by tuning $\alpha_{\rm R}$.

\section{Summary}\label{sec:sum}
We have theoretically studied the magnetization reversal by tuning the RSOI ($\alpha_{\rm R}$) and Josephson phase in an {\it S/N/F/S} junction. 
The magnetizations of the {\it x} and {\it z} components are induced by the appearance of the odd-frequency spin-triplet 
and even-frequency spin-singlet Cooper pairs in the {\it N}. 
We have shown that the magnetizations exhibit damped oscillatory behavior as a function of the thickness of $N$ 
for finite $\alpha_{\rm R}$. 
The period of oscillation of the magnetizations induced by the proximity effect is varied by changing $\alpha_{\rm R}$ and 
becomes short with increasing $\alpha_{\rm R}$. 
Therefore, the direction of magnetizations can be controlled by tuning $\alpha_{\rm R}$ for a fixed thickness of {\it N}. 
We have found that the magnetizations induced in the {\it N} depend on the Josephson phase ($\theta$). 
As a result, the amplitude and the oscillation period of the magnetizations can be controlled by tuning $\theta$. 
It has also been found that the direction of the magnetizations in the {\it N} can be reversed by changing $\theta$ as well as $\alpha_{\rm R}$. 
These results clearly show that the variation of the magnetization by tuning $\alpha_{\rm R}$ and $\theta$ is a good means of observing the spin of STCs. 

We have theoretically shown that the magnetizations are decomposed into three parts, i.e., $M_{x(z)}(d,\theta)=M^{(1)}_{x(z)}(d)+M^{(2)}_{x(z)}(d,\theta)+M^{(3)}_{x(z)}(d,\theta)$. 
(i) The appearance of $M^{(1)}_{x}(d)$, which is the $\theta$-independent part of the magnetization, is due to the proximity effect, the exchange field in the {\it F},  
and the RSOI in the {\it N}. 
On the other hand, the $\theta$-dependent parts $M^{(2)}_{x}(d,\theta)$ and $M^{(3)}_{x}(d,\theta)$ 
result from the finite Josephson coupling between the two superconductors, the exchange field in the {\it F}, and the RSOI in the {\it N} in the {\it S/N/F/S} junction. 
(ii) $M_{z}^{(1)}(d)$, which always appears in the {\it S/F} junctions, is induced by the proximity effect. 
The $\theta$-dependent part $M_{z}^{(2)}(d,\theta)$ results from the finite Josephson coupling between two superconductors and the exchange field in the {\it F}. 
$M_{z}^{(3)}(d,\theta)$ appears only when the Josephson coupling, the exchange field, and the RSOI are finite. 
For the $\theta$-dependence of the magnetizations, 
we have found that $M_{x(z)}^{(2)}(d,\theta)$ is a cosine function of $\theta$ and $M_{x(z)}^{(3)}(d,\theta)$ is sinusoidal with $\theta$. 

We have also shown that the magnetization induced by the proximity effect can be large enough to be detected in typical experiments. 
Therefore, it is expected that a Josephson junction including the {\it F} and the RSOI in the {\it N}, 
such as the one studied here, has a potential for low-Joule-heating spintronics devices, 
since the direction of the magnetization inside the {\it N} can be easily controlled by changing $\alpha_{\rm R}$ and $\theta$. 

\section*{ACKNOWLEDGMENTS} 
The authors would like to thank M. Mori for useful discussions and comments. 

\appendix{

\section{Coefficients $C^{ij}_{\omega_{n}}(d)$}
In Eqs.~(\ref{phiw}) and (\ref{txw}), the coefficients $C^{ij}_{\omega_{n}}(d)$ are given by
%
\begin{eqnarray}
C_{{\omega _n}}^{21}\left( d \right) &=& 
i{\alpha _{\rm{R}}}{d_{\rm{F}}}\left[ {\frac{{\left( {i{\alpha _{\rm{R}}} - {k_\alpha }} \right){d_{\rm{F}}}}}{{{\gamma _{\rm{F}}}}}{e^{\left( { - i{\alpha _{\rm{R}}} + {k_\alpha }} \right)L}} 
- \frac{{\left( {i{\alpha _{\rm{R}}} + {k_\alpha }} \right){d_{\rm{F}}}}}{{{\gamma _{\rm{F}}}}}{e^{ - \left( {i{\alpha _{\rm{R}}} + {k_\alpha }} \right)L}}} \right] \nonumber \\
&+& {k_\alpha }{d_{\rm{F}}}\left[ {\frac{{\left( {i{\alpha _{\rm{R}}} - {k_\alpha }} \right){d_{\rm{F}}}}}{{{\gamma _{\rm{F}}}}}{e^{\left( {i{\alpha _{\rm{R}}} 
- {k_\alpha }} \right)L}} + \frac{{\left( {i{\alpha _{\rm{R}}} + {k_\alpha }} \right){d_{\rm{F}}}}}{{{\gamma _{\rm{F}}}}}{e^{ - \left( {i{\alpha _{\rm{R}}} + {k_\alpha }} \right)L}}} \right], \\
C_{{\omega _n}}^{22}\left( d \right) 
&=&  - \left( {i{\alpha _{\rm{R}}} + {k_\alpha }} \right)
{d_{\rm{F}}}\left[ {\frac{{\left( {i{\alpha _{\rm{R}}} - {k_\alpha }} \right){d_{\rm{F}}}}}{{{\gamma _{\rm{F}}}}}{e^{\left( { - i{\alpha _{\rm{R}}} + {k_\alpha }} \right)L}} 
- \frac{{\left( {i{\alpha _{\rm{R}}} + {k_\alpha }} \right){d_{\rm{F}}}}}{{{\gamma _{\rm{F}}}}}{e^{ - \left( {i{\alpha _{\rm{R}}} + {k_\alpha }} \right)L}}} \right] \nonumber \\
&-& {k_\alpha }{d_{\rm{F}}}\frac{{\left( {i{\alpha _{\rm{R}}} + {k_\alpha }} \right){d_{\rm{F}}}}}{{{\gamma _{\rm{F}}}}}\left[ {{e^{\left( {i{\alpha _{\rm{R}}} + {k_\alpha }} \right)L}} 
+ {e^{ - \left( {i{\alpha _{\rm{R}}} + {k_\alpha }} \right)L}}} \right], \\
C_{{\omega _n}}^{23}\left( d \right) 
&=&  - \left( {i{\alpha _{\rm{R}}} + {k_\alpha }} \right){d_{\rm{F}}}\left[ {\frac{{\left( {i{\alpha _{\rm{R}}} 
- {k_\alpha }} \right){d_{\rm{F}}}}}{{{\gamma _{\rm{F}}}}}{e^{\left( {i{\alpha _{\rm{R}}} - {k_\alpha }} \right)L}} + \frac{{\left( {i{\alpha _{\rm{R}}} 
+ {k_\alpha }} \right){d_{\rm{F}}}}}{{{\gamma _{\rm{F}}}}}{e^{ - \left( {i{\alpha _{\rm{R}}} + {k_\alpha }} \right)L}}} \right] \nonumber \\
&+& i{\alpha _{\rm{R}}}{d_{\rm{F}}}\frac{{\left( {i{\alpha _{\rm{R}}} + {k_\alpha }} \right){d_{\rm{F}}}}}{{{\gamma _{\rm{F}}}}}
\left[ {{e^{\left( {i{\alpha _{\rm{R}}} + {k_\alpha }} \right)L}} + {e^{ - \left( {i{\alpha _{\rm{R}}} + {k_\alpha }} \right)L}}} \right], \\
C_{{\omega _n}}^{31}\left( d \right) 
&=& i{\alpha _{\rm{R}}}{d_{\rm{F}}}
\left\{ {\left[ {1 - \frac{{\left( {i{\alpha _{\rm{R}}} 
- {k_\alpha }} \right){d_{\rm{F}}}}}{{{\gamma _{\rm{F}}}}}} \right]{e^{{k_\alpha }d}} 
- \left[ {1 + \frac{{\left( {i{\alpha _{\rm{R}}} + {k_\alpha }} \right){d_{\rm{F}}}}}{{{\gamma _{\rm{F}}}}}} \right]
{e^{ - {k_\alpha }d}}} \right\}{e^{ - i\tilde \alpha d}} \nonumber \\
&-& {k_\alpha }{d_{\rm{F}}}
\left\{ {\left[ {1 + \frac{{\left( {i{\alpha _{\rm{R}}} - {k_\alpha }} \right){d_{\rm{F}}}}}{{{\gamma _{\rm{F}}}}}} \right]{e^{i{\alpha _{\rm{R}}}d}} 
- \left[ {1 + \frac{{\left( {i{\alpha _{\rm{R}}} + {k_\alpha }} \right){d_{\rm{F}}}}}{{{\gamma _{\rm{F}}}}}} \right]{e^{ - i\tilde \alpha d}}} \right\}{e^{ - {k_\alpha }d}}, \\
C_{{\omega _n}}^{32}\left( d \right) 
&=&  - \left( {i{\alpha _{\rm{R}}} + {k_\alpha }} \right){d_{\rm{F}}}
\left\{ {\left[ {1 - \frac{{\left( {i{\alpha _{\rm{R}}} - {k_\alpha }} \right){d_{\rm{F}}}}}{{{\gamma _F}}}} \right]{e^{{k_\alpha }d}} 
- \left[ {1 + \frac{{\left( {i{\alpha _{\rm{R}}} + {k_\alpha }} \right){d_{\rm{F}}}}}{{{\gamma _F}}}} \right]{e^{ - {k_\alpha }d}}} \right\}{e^{ - i{\alpha _{\rm{R}}}d}} \nonumber \\
&+& {k_\alpha }{d_{\rm{F}}}\left[ {1 + \frac{{\left( {i{\alpha _{\rm{R}}} + {k_\alpha }} \right){d_{\rm{F}}}}}{{{\gamma _F}}}} \right]\left[ {{e^{\left( {i{\alpha _{\rm{R}}} 
+ {k_\alpha }} \right)d}} - {e^{ - \left( {i{\alpha _{\rm{R}}} + {k_\alpha }} \right)d}}} \right], 
\end{eqnarray}
%
and
%
\begin{eqnarray}
C_{{\omega _n}}^{33}\left( d \right) &=& 
	\left( {i{\alpha _{\rm{R}}} + {k_\alpha }} \right){d_{\rm{F}}}\left\{ {\left[ {1 + \frac{{\left( {i{\alpha _{\rm{R}}} - {k_\alpha }} \right)
	{d_{\rm{F}}}}}{{{\gamma _F}}}} \right]{e^{i\tilde \alpha d}} - \left[ {1 + \frac{{\left( {i{\alpha _{\rm{R}}} + {k_\alpha }} \right){d_{\rm{F}}}}}{{{\gamma _F}}}} \right]
	{e^{ - i{\alpha _{\rm{R}}}d}}} \right\}{e^{ - {k_\alpha }d}} \nonumber \\
	&-& i{\alpha _{\rm{R}}}{d_{\rm{F}}}\left[ {1 + \frac{{\left( {i{\alpha _{\rm{R}}} + {k_\alpha }} \right){d_{\rm{F}}}}}{{{\gamma _F}}}} \right]
	\left[ {{e^{\left( {i{\alpha _{\rm{R}}} + {k_\alpha }} \right)d}} - {e^{ - \left( {i{\alpha _{\rm{R}}} + {k_\alpha }} \right)d}}} \right].
\end{eqnarray}
%
%
%
\section{Integration with respect to $x$ in Eq.~(\ref{md})}
In this Appendix, we will first provide the analytical form of the local magnetization density in the $N$. 
Within quasiclassical Green's function theory, the local magnetization density ${\bm m}(x,\theta)$ in the $N$ 
is obtained by substituting Eqs.~(\ref{fns})--(\ref{fntz}) into Eq.~({\ref{lm}}). 
The {\it x} component $m_{x}(x,\theta)$ of the local magnetization density can be decompose into 
$\theta$-independent and $\theta$-dependent parts,
\begin{eqnarray}
m_{x}(x,\theta) &=& m_x^{\left( 1 \right)}\left( x \right) + m_x^{\left( 2 \right)}\left( {x,\theta } \right) + m_x^{\left( 3 \right)}\left( {x,\theta } \right),
\label{lmx-apb}
\end{eqnarray}
where 
\begin{eqnarray}
m_{x}^{(1)}(x,\theta) &=&
-g \mu_{\rm B} \pi N_{\rm F} k_{\rm B} T
\frac{h_{\rm ex} d_{\rm F}^{2}}{\hbar D_{\rm F}} 
\sum_{i \omega_{n}}
\frac{\Delta^{2}}{E_{\omega_{n}}^{2}} 
Q_{\omega_{n}}(d)  \nonumber \\
&\times &
\sinh(k_{\rm N} x)
{\rm Im}
\left[
\Phi_{\omega_{n}}(d) t_{\omega_{n}}(x,d)
\right]
\label{lmx1}, \\
m_{x}^{(2)}(x,\theta) &=&
g \mu_{\rm B} \pi N_{\rm F} k_{\rm B} T
\frac{h_{\rm ex} d_{\rm F}^{2}}{\hbar D_{\rm F}} 
\sum_{i \omega_{n}}
\frac{\Delta^{2}}{E_{\omega_{n}}^{2}} 
\left(
1-\frac{k_{\rm F}d_{\rm F}}{\gamma_{\rm F}}
\right) \nonumber \\
&\times &
Q_{\omega_{n}}(d)  
\sinh[k_{\rm N} (x-d)] \nonumber \\
&\times &
{\rm Im}
\left[
\Phi_{\omega_{n}}(d) t_{\omega_{n}}(x,d)
\right] \cos \theta
\label{lmx2}, \nonumber \\
\end{eqnarray}
and
\begin{eqnarray}
m_{x}^{(3)}(x,\theta) &=&
g \mu_{\rm B} \pi N_{\rm F} k_{\rm B} T
\frac{h_{\rm ex} d_{\rm F}^{2}}{\hbar D_{\rm F}} 
\sum_{i \omega_{n}}
\frac{\Delta^{2}}{E_{\omega_{n}}^{2}} 
\left(
1-\frac{k_{\rm F}d_{\rm F}}{\gamma_{\rm F}}
\right) \nonumber \\
&\times& Q_{\omega_{n}}(d)
\sinh[k_{\rm N} (x-d)] \nonumber \\
&\times &{\rm Re}
\left[
\Phi_{\omega_{n}}(d) t_{\omega_{n}}(x,d)
\right] \sin \theta
\label{lmx3}, 
\end{eqnarray}
where the functions $Q_{\omega_{n}}(d)$, $\Phi_{\omega_{n}}(d)$, and $t_{\omega_{n}}(x,d)$ are respectively given in Eqs.~(\ref{qwn})--(\ref{txw}). 
Substituting Eqs.~(\ref{lmx1})--(\ref{lmx3}) into Eq.~(\ref{md}) and performing the integration with respect to $x$, 
we can obtain the $x$ component of the magnetization, 
%
\begin{eqnarray}
\begin{array}{l}
{M_x}\left( {d,\theta } \right) = \frac{1}{d}\int_0^d {{m_x}\left( {x,\theta } \right)dx} = \frac{1}{d}\int_0^d {m_x^{\left( 1 \right)}\left( x \right)dx}  
+ \frac{1}{d}\int_0^d {m_x^{\left( 2 \right)}\left( {x,\theta } \right)dx}  + \frac{1}{d}\int_0^d {m_x^{\left( 3 \right)}\left( {x,\theta } \right)dx} \nonumber \\
 =  - g{\mu _{\rm{B}}}\pi {N_{\rm{F}}}k_{\rm B}T\frac{1}{d}\sum\limits_{i{\omega _n}} {\frac{{{h_{{\rm{ex}}}}d_{\rm{F}}^2}}{{\hbar {D_{\rm{F}}}}}
\frac{{{\Delta ^2}}}{{E_{{\omega _n}}^2}}{Q_{{\omega _n}}}\left( d \right)\int_0^d {\sinh \left( {{k_{\rm{N}}}x} \right){\mathop{\rm Im}\nolimits} 
\left[ {{\Phi _{{\omega _n}}}\left( d \right){t_{{\omega _n}}}\left( {x,d} \right)} \right]dx} } \nonumber \\
 + g{\mu _{\rm{B}}}\pi {N_{\rm{F}}}k_{\rm B}T\frac{1}{d}\sum\limits_{i{\omega _n}} {\frac{{{h_{{\rm{ex}}}}d_{\rm{F}}^2}}{{\hbar {D_{\rm{F}}}}}
\frac{{{\Delta ^2}}}{{E_{{\omega _n}}^2}}\left( {1 - \frac{{{k_{\rm{N}}}{d_{\rm{F}}}}}{{{\gamma _{\rm{F}}}}}} \right){Q_{{\omega _n}}}\left( d \right)\int_0^d 
{\sinh \left[ {{k_{\rm{N}}}\left( {x - d} \right)} \right]{\mathop{\rm Im}\nolimits} \left[ {{\Phi _{{\omega _n}}}\left( d \right){t_{{\omega _n}}}\left( {x,d} \right)} \right]dx} \cos \theta } \nonumber \\
 + g{\mu _{\rm{B}}}\pi {N_{\rm{F}}}k_{\rm B}T\frac{1}{d}\sum\limits_{i{\omega _n}} {\frac{{{h_{{\rm{ex}}}}d_{\rm{F}}^2}}{{\hbar {D_{\rm{F}}}}}\frac{{{\Delta ^2}}}{{E_{{\omega _n}}^2}}\left( {1 - \frac{{{k_{\rm{N}}}{d_{\rm{F}}}}}{{{\gamma _{\rm{F}}}}}} \right){Q_{{\omega _n}}}\left( d \right)\int_0^d {\sinh 
\left[ {{k_{\rm{N}}}\left( {x - d} \right)} \right]{\mathop{\rm Re}\nolimits} \left[ {{\Phi _{{\omega _n}}}\left( d \right){t_{{\omega _n}}}
\left( {x,d} \right)} \right]dx} \sin \theta } \nonumber \\
 = M_x^{\left( 1 \right)}\left( {d,\theta } \right) + M_x^{\left( 2 \right)}\left( {d,\theta } \right) + M_x^{\left( 3 \right)}\left( {d,\theta } \right), \nonumber
\end{array}
\end{eqnarray}
where $M_{x}^{(1)}(d)$, $M_{x}^{(2)}(d,\theta)$, and $M_{x}^{(3)}(d,\theta)$ are respectively given in Eqs.~(\ref{mxd1})--(\ref{mxd3}). 
The functions $w_{\omega_{n}}(d)$ and $u_{\omega_{n}}(d)$ in Eqs.~(\ref{mxd1})--(\ref{mxd3}) are given as 
%
\begin{eqnarray}
u_{\omega_{n}}(d) &=& u_{\omega_{n}}^{(1)}(d) + u_{\omega_{n}}^{(2))}(d) + u_{\omega_{n}}^{(3)}(d), \\
u_{{\omega _n}}^{\left( 1 \right)}\left( d \right) &=&  
- 2\left[ {C_{{\omega _n}}^{21}\left( d \right) + 2C_{{\omega _n}}^{31}\left( d \right)} \right]
\frac{{{k_{\rm{N}}}\sin \left[ {\left( {{\alpha _{\rm{R}}} - i{k_\alpha }} \right)d} \right] 
- \left( {{\alpha _{\rm{R}}} - i{k_\alpha }} \right)
\sinh \left( {{k_{\rm{N}}}d} \right)}}{{{{\left( {{\alpha _{\rm{R}}} - i{k_\alpha }} \right)}^2} + k_{\rm{N}}^2}}, \nonumber \\
u_{{\omega _n}}^{\left( 2 \right)}\left( d \right) &=& 
4{\alpha _{\rm{R}}}\left[ {C_{{\omega _n}}^{22}\left( d \right) + 2C_{{\omega _n}}^{32}\left( d \right)} \right]{e^{ - {k_\alpha }d}} \nonumber \\
&\times &
\frac{{{k_\alpha }{k_{\rm{N}}}
\cos \left( {{\alpha _{\rm{R}}}d} \right) + {k_{\rm{N}}}{\alpha _{\rm{R}}}
\sin \left( {{\alpha _{\rm{R}}}d} \right) - {e^{{k_\alpha }d}}\left[ {{k_\alpha }{k_{\rm{N}}}
\cosh \left( {{k_{\rm{N}}}d} \right) - \left( {2\alpha _{\rm{R}}^2 + k_{\rm{N}}^2} \right)
\sinh \left( {{k_{\rm{N}}}d} \right)} \right]}}{{{{\left( {\alpha _{\rm{R}}^2 + k_\alpha ^2} \right)}^2} 
+ 2\left( {\tilde \alpha _{\rm{R}}^2 - k_\alpha ^2} \right)k_{\rm{N}}^2 + k_{\rm{N}}^4}}, \nonumber \\
u_{{\omega _n}}^{\left( 3 \right)}\left( d \right) &=& 
- i4{\alpha _{\rm{R}}}\left[ {C_{{\omega _n}}^{23}\left( d \right) + 2C_{{\omega _n}}^{33}\left( d \right)} \right] \nonumber \\
&\times &
\frac{{i{k_\alpha }{k_{\rm{N}}}\cosh \left( {{k_\alpha }d} \right) + {k_{\rm{N}}}{\alpha _{\rm{R}}}
\sinh \left( {{k_\alpha }d} \right) - {\alpha _{\rm{R}}}{k_\alpha }\left[ {i{k_{\rm{N}}}\cosh \left( {{k_{\rm{N}}}d} \right) + 2{\alpha _{\rm{R}}}
\sinh \left( {{k_{\rm{N}}}d} \right)} \right]{e^{i{\alpha _{\rm{R}}}d}}}}{{\left[ {\alpha _{\rm{R}}^2 + {{\left( {{k_\alpha } - {k_{\rm{N}}}} \right)}^2}} \right]
\left[ {\alpha _{\rm{R}}^2 + {{\left( {{k_\alpha } + {k_{\rm{N}}}} \right)}^2}} \right]}}{e^{ - i{\alpha _{\rm{R}}}d}} \nonumber
\end{eqnarray}
%
\begin{eqnarray}
w_{\omega_{n}}(d) &=& w_{\omega_{n}}^{(1)}(d) + w_{\omega_{n}}^{(2))}(d) + w_{\omega_{n}}^{(3)}(d), \\
w_{{\omega _n}}^{\left( 1 \right)}\left( d \right) &=& 
\frac{{{e^{ - \left( {i{\alpha _{\rm{R}}} + {k_\alpha }} \right)d}}}}{{{{\left( {{\alpha _{\rm{R}}} - i{k_\alpha }} \right)}^2} 
+ k_{\rm{N}}^2}}\left[ {\left( { - 1 + {e^{\left( {i{\alpha _{\rm{R}}} + {k_\alpha }} \right)2d}}} \right){k_{\rm{N}}}\cosh \left( {{k_{\rm{N}}}d} \right) 
- i\left( {1 + {e^{\left( {i{\alpha _{\rm{R}}} + {k_\alpha }} \right)2d}}} \right)\left( {{\alpha _{\rm{R}}} - i{k_\alpha }} \right)
\sinh \left( {{k_\alpha }d} \right)} \right], \nonumber \\
w_{{\omega _n}}^{\left( 2 \right)}\left( d \right) &=& 
{e^{ - {k_\alpha }d}}\frac{{a_{{\omega _n}}^{\left( 1 \right)}\left( d \right) + 
a_{{\omega _n}}^{\left( 2 \right)}\left( d \right) - a_{{\omega _n}}^{\left( 3 \right)}\left( d \right)}}{{{{\left( {\alpha _{\rm{R}}^2 + k_\alpha ^2} \right)}^2} 
- \left( {4\alpha _{\rm{R}}^2 - k_{\rm{N}}^2} \right)k_{\rm{N}}^2}}, \nonumber \\
a_{\omega_{n}}^{(1)}(d) &=& 2 \alpha_{\rm R} k_{\alpha} k_{\rm N} e^{k_{\alpha} d}, \nonumber \\
a_{{\omega _n}}^{\left( 2 \right)}\left( d \right) &=&  
- 2{\alpha _{\rm{R}}}{k_{\rm{N}}}\cosh \left( {{k_{\rm{N}}}d} \right)\left[ {{k_\alpha }\cos \left( {{\alpha _{\rm{R}}}d} \right) 
+ \tilde \alpha \sin \left( {{\alpha _{\rm{R}}}d} \right)} \right], \nonumber \\
a_{{\omega _n}}^{\left( 3 \right)}\left( d \right) &=& 
2{\alpha _{\rm{R}}}\left[ {\left( {2\alpha _{\rm{R}}^2 + k_{\rm{N}}^2} \right)
\cos \left( {{\alpha _{\rm{R}}}d} \right) + 2{k_\alpha }{\alpha _{\rm{R}}}
\sin \left( {{\alpha _{\rm{R}}}d} \right)} \right] \sinh \left( {{k_{\rm{N}}}d} \right), \nonumber \\
w_{{\omega _n}}^{\left( 3 \right)}\left( d \right) &=& 
\frac{1}{2}{e^{ - i{\alpha _{\rm{R}}}d}}\left[ {b_{{\omega _n}}^{\left( 1 \right)}\left( d \right) 
+ b_{{\omega _n}}^{\left( 2 \right)}\left( d \right) + b_{{\omega _n}}^{\left( 3 \right)}\left( d \right)} \right], \\
b_{{\omega _n}}^{\left( 1 \right)}\left( d \right) &=& 
{e^{i{\alpha _{\rm{R}}}d}}\left[ {\frac{{i{\alpha _{\rm{R}}}}}{{\alpha _{\rm{R}}^2 + {{\left( {{k_\alpha } - {k_{\rm{N}}}} \right)}^2}}} 
- \frac{{i{\alpha _{\rm{R}}}}}{{\alpha _{\rm{R}}^2 + {{\left( {{k_\alpha } + {k_{\rm{N}}}} \right)}^2}}}} \right], \nonumber \\
b_{{\omega _n}}^{\left( 2 \right)}\left( d \right) &=&  
- \frac{{i{\alpha _{\rm{R}}}\cosh \left[ {\left( {{k_\alpha } - {k_{\rm{N}}}} \right)d} \right] 
+ \left( {{k_\alpha } - {k_{\rm{N}}}} \right)\sinh \left[ {\left( {{k_\alpha } - {k_{\rm{N}}}} \right)d} \right]}}{{\alpha _{\rm{R}}^2 
+ {{\left( {{k_\alpha } - {k_{\rm{N}}}} \right)}^2}}}, \nonumber 
\end{eqnarray}
%
and
%
\begin{eqnarray}
b_{{\omega _n}}^{\left( 3 \right)}\left( d \right) &=& 
\frac{{i{\alpha _{\rm{R}}}\cosh \left[ {\left( {{k_\alpha } + {k_{\rm{N}}}} \right)d} \right] 
+ \left( {{k_\alpha } + {k_{\rm{N}}}} \right)\sinh \left[ {\left( {{k_\alpha } + {k_{\rm{N}}}} \right)d} \right]}}
{{\alpha _{\rm{R}}^2 + {{\left( {{k_\alpha } + {k_{\rm{N}}}} \right)}^2}}} \nonumber.
\end{eqnarray}
%

We can also obtain the $z$ component of the magnetization given in Eqs.~(\ref{mzd1})--(\ref{mzd3}) 
by following the procedure used to derive the $x$ component of the magnetization. 

}
{}

\begin{thebibliography}{}
\bibitem{note1}
In the {\it S/N/F/S} junction without the RSOI, the STC composed of opposite spin electrons 
is only induced by the proximity effect inside the $N$, and thus only the {\it z} component of magnetization is induced\cite{bergeret-rmp}.
\bibitem{note}
We need to apply the DC bias current to the present junction to change $\theta$. 
Typically, the magnitude of the bias current is about 10 $\mu $A and the size of the device is about 1 $\mu $m\cite{frolov-prb}. 
In this case, the magnitude of the magnetic field ({\it H}) induced by the DC bias current is about 20 A/m based on Ampere's law $H=I/2 \pi r$, 
where $I$ and $r$ are the DC bias current and the distance from circuit, respectively. 
This $H$ is much smaller than the magnetization induced by the proximity effect. 
Therefore, the influence of {\it H} can be neglected.
\bibitem{note7}
$N_{\rm F}=\frac{1}{4\pi^2}\left( \frac{2m}{\hbar^2} \right)^{3/2}\varepsilon_{\rm F}^{1/2} \approx 5.4 \times10^{27}\,\ {\rm eV^{-1}} {\rm m^{-3}} $ 
per spin with the Fermi energy 
$\varepsilon_{\rm F}\approx5$ eV~\cite{book-ssp}, where $m$ is the electron mass. 
$\Delta_{0}=1$ meV for Nb~\cite{book-deutscher}.
\end{thebibliography}
\end{document}